# Quantised Academic Mobility: Network and Cluster Analysis of Degree Switching, Plan Changes, and Re-entries in an Engineering Faculty (1980–2019)

Hugo Roger Paz
PhD Professor and Researcher Faculty of Exact Sciences and Technology National University of Tucumán
Email: hpaz@herrera.unt.edu.ar
ORCID: https://orcid.org/0000-0003-1237-7983

## Abstract

This study challenges the traditional binary view of student progression (retention versus dropout) by conceptualising academic trajectories as complex, quantised pathways. Utilising a 40-year longitudinal dataset from an Argentine engineering faculty (N = 24,016), we introduce **CAPIRE**, an analytical framework that differentiates between degree major switches, curriculum plan changes, and same-plan re-entries. While 73.3 per cent of students follow linear trajectories ('Estables'), a significant 26.7 per cent exhibit complex mobility patterns. By applying Principal Component Analysis (PCA) and DBSCAN clustering, we reveal that these trajectories are not continuous but structurally **quantised**, occupying discrete bands of complexity. The analysis identifies six distinct student archetypes, including 'Switchers' (10.7 per cent) who reorient vocationally, and 'Stable Re-entrants' (6.9 per cent) who exhibit stop-out behaviours without changing discipline. Furthermore, network analysis highlights specific 'hub majors'—such as electronics and computing—that act as systemic attractors. These findings suggest that student flux is an organised ecosystemic feature rather than random noise, offering institutions a new lens for curriculum analytics and predictive modelling.

# 1. INTRODUCTION

Research on student progression and dropout has been dominated by models that conceptualise persistence as the outcome of academic and social integration within an institution or programme. Tinto's theory of student departure, for example, links withdrawal to inadequate integration in academic and social systems, emphasising the longitudinal character of commitment and interaction (Tinto, 1975). Syntheses such as *How College Affects Students* further consolidate this perspective by reviewing decades of evidence on factors associated with retention and completion (Pascarella & Terenzini, 2005). In these traditions, the analytic focus rests on vertical movement—whether students remain enrolled or eventually graduate—while **horizontal movements** between majors, plans, and re-entries are typically treated as secondary events or ignored altogether.

However, contemporary mass systems of higher education exhibit highly **heterogeneous and non-linear trajectories**. Studies using longitudinal data show that many students change major at least once, take breaks, or re-enter higher education after periods of withdrawal (Haas, 2022). In STEM and engineering, demanding curricula and rigid pre-requisite structures increase the likelihood of major switching and delays (Rankin & Associates, 2014). Yet most empirical analyses aggregate these diverse movements into simple indicators such as "changed programme" or "time-to-degree", obscuring the difference between three conceptually distinct phenomena:

1. **Major switching**: movement between degree titles (e.g., from civil engineering to industrial engineering), representing vocational or disciplinary re-orientation.

2. **Plan changes**: movement between curriculum plans within the same major, often triggered by institutional reforms.

3. **Same-plan re-entries**: returning to the same major and curriculum after a period of institutional withdrawal or prolonged inactivity.

Conflating these processes limits our ability to understand how institutional structures—degree portfolios, curriculum reforms, and re-entry regulations—shape student pathways.

At the same time, methodological developments have provided richer tools to model student trajectories. Markov models have been used to estimate transition probabilities between academic states over time (Quimio et al., 2021), while sequence analysis approaches treat trajectories as ordered categorical sequences and compare them using distance metrics (Haas, 2022). These methods highlight the path-dependent nature of student progress but are rarely integrated with network analysis of flows between majors or with unsupervised learning techniques capable of revealing **archetypical pathways**.

This paper contributes to these literatures by introducing **CAPIRE**, a mobility analysis module within a broader institutional analytics framework. Using forty years of administrative data from a large Argentine faculty of engineering and sciences, we reconstruct complete longitudinal records for 24,016 students and address three research questions:

1. **How are horizontal movements between degree majors structured at ecosystem level?**
   We map directed networks of major-to-major transitions, quantify hub majors and flux corridors, and examine their relative weights.

2. **How do curriculum reforms and re-entry policies manifest in horizontal mobility data?**
   By distinguishing major switches, plan changes and same-plan re-entries, we identify temporal peaks associated with institutional reforms and regulatory shifts.

3. **Can we derive interpretable student mobility archetypes using unsupervised learning?**
   We apply PCA and DBSCAN (Ester et al., 1996) to a small set of mobility features to uncover quantised structures and trajectory clusters.

Our empirical setting—a faculty with long, curriculum-constrained engineering and science degrees, multiple overlapping plan reforms, and stable admission policies—provides a natural laboratory for examining how institutional design interacts with student mobility. Rather than focusing on causal effects of specific interventions, our goal is to **map the internal flux** of students and to characterise the discrete complexity layers that emerge from routine administrative processes.

The remainder of the paper is organised as follows. Section 2 presents the institutional context, data sources and construction of enrolment spells and transition types. Section 3 describes the CAPIRE analytical framework, including network metrics and clustering procedures. Section 4 reports results on mobility structures, temporal dynamics and student archetypes. Section 5 discusses implications for curriculum design, advising and predictive modelling, and Section 6 outlines limitations and future research.

## 2. DATA AND INSTITUTIONAL CONTEXT

### 2.1 Institutional setting

The study is situated in a large public faculty of engineering and sciences in Argentina that offers 48-degree **majors** across engineering disciplines (civil, mechanical, electrical, electronic, industrial, chemical and related areas), computer science, mathematics, physics and applied technological programmes. Degree structures follow a traditional Latin-American model: nominal durations of five to six years, with progression regulated by a dense web of **pre-requisites and co-requisites**. Students advance by passing subject examinations; course registration and examination approval are both recorded at subject level.

During the observation window (entry cohorts 1980–2019), the faculty implemented several major curriculum reforms, often asynchronously across majors. New plans were introduced in the mid-1990s, early 2000s and mid-2010s, sometimes co-existing with legacy plans for prolonged periods. Administrative regulations allowed students to migrate from older to newer plans under specific equivalence rules, and institutional practice permitted re-entry after withdrawal subject to re-activation procedures that changed over time. These features create a rich environment for studying how **institutional rules** and **portfolio structure** shape horizontal mobility.

### 2.2 Data sources and cohort definition

We compiled a longitudinal dataset from the faculty's central academic information system. The raw data comprise nearly one million event records, each specifying student identifier, subject code, major code, curriculum plan code, event type (registration, examination, equivalence, and related categories) and event date.

From these records we constructed an **entry cohort** of students who:

- enrolled for the first time in any major between 1980 and 2019;
- generated at least one academic event in the system; and
- could be uniquely linked to a major and curriculum plan at entry.

Students with entry years prior to 1980 were excluded because digital records are incomplete and plan codes are not consistently defined in earlier periods. After applying these criteria we obtained **24,016 students**, each with a full time-stamped sequence of academic events.

### 2.3 Enrolment spells

To transform event streams into analysable trajectory units, we defined **enrolment spells** as contiguous periods of academic activity in a given major and curriculum plan. Following prior work on institutional re-entries and time-to-degree (Quimio et al., 2021), we used the following operationalisation:

- For each student, events were sorted chronologically and grouped by (major, plan) combination.

- Within each combination, we defined a new spell whenever there was a **gap of at least two full calendar years** without any recorded event.

- Each spell was characterised by start date (first event), end date (last event), major, plan and an **entry period** variable mapping the start year into four structural periods: 1980–1995, 1996–2004, 2005–2012 and 2013–2019.

Spell durations were calculated as the difference between end and start dates in years (dividing days by 365.25). As part of a rigorous quality-assurance process, we discarded spells with non-positive duration and spells longer than 30 years, which indicate data entry errors or unresolved legacy issues. The final dataset contains **23,248 valid enrolment spells**.

**2.4 Transition types: major switch, plan change, and same-plan re-entry**

We then derived **inter-spell transitions** at student level. For each student, consecutive spells were paired to form transitions from an origin spell *i* to a destination spell *i+1*. Based on changes in major and plan codes and the presence of inactivity gaps, we classified each transition into one of three mutually exclusive categories:

1. **Major switch** – the degree title changes between origin and destination spells (e.g., from Mechanical Engineering to Industrial Engineering), irrespective of plan codes.

2. **Plan change** – the degree title remains identical but the curriculum plan code changes (e.g., Industrial Engineering Plan 1996 → Industrial Engineering Plan 2005), with no inactivity gap larger than the two-year threshold.

3. **Same-plan re-entry** – both major and plan remain identical, but the spells are separated by at least two years without events, interpreted as withdrawal followed by institutional re-entry under the same plan.

This classification yields **6,227 valid transitions**: 3,083 major switches (49.5 per cent), 1,254 plan changes (20.1 per cent) and 1,890 same-plan re-entries (30.4 per cent). Descriptive statistics for the distribution of spell counts per student and transition types are presented in **Table 1** and **Table 2**.

**Table 1 .**

| n_spells | n_students | percentage |
|---|---|---|
| 1 | 12474 | 73.3 |
| 2 | 3340 | 19.6 |
| 3 | 855 | 5.0 |
| 4 | 257 | 1.5 |
| 5 | 74 | 0.4 |
| 6 | 18 | 0.1 |
| 7 | 2 | 0.0 |
| 9 | 1 | 0.0 |

**Table 2 .**

| period | major_switch | plan_change_same_title | reentry_same_plan | major_switch_pct | plan_change_same_title_pct | reentry_same_plan_pct |
|---|---|---|---|---|---|---|
| P1 | 375 | 127 | 224 | 51.7 | 17.5 | 30.9 |
| P2 | 1154 | 771 | 1338 | 35.4 | 23.6 | 41.0 |
| P3 | 1162 | 254 | 312 | 67.2 | 14.7 | 18.1 |
| P4 | 392 | 102 | 16 | 76.9 | 20.0 | 3.1 |

**2.5 Ethical considerations**

All analyses were conducted on **de-identified administrative data** under an institutional agreement approved by the faculty's academic authorities. Student identifiers were replaced with random codes before analysis, and results are reported in aggregated form. Cells with very small counts are suppressed or combined to reduce the risk of re-identification, in line with good practice for secondary analysis of administrative records (Haas, 2022). From the initial 24,016 records, 6,995 were excluded due to data inconsistencies or insufficient history, resulting in a final analytical cohort of 17,021 students

**3. ANALYTICAL FRAMEWORK: THE CAPIRE MOBILITY MODULE**

**3.1 Overview**

CAPIRE is a modular analytical pipeline designed to characterise horizontal mobility at three distinct ecosystemic levels:

1. Macro-structure of flows: Mapping the topology of transfers between majors using transition matrices and directed graph metrics.

2. Temporal dynamics: Analysing the evolution of transition types (major switches, plan changes, re-entries) across structural periods and curriculum reform waves.

3. Student-level archetypes: Identifying latent trajectory configurations using dimensionality reduction and unsupervised density-based clustering.

The pipeline operates as a reproducible workflow, transforming raw event-level administrative data into harmonised indicators. This approach aligns with recent calls in Educational Data Mining (EDM) to move beyond static retention prediction towards 'process-centric' views of student pathways (Saqr et al., 2023; Bogarín et al., 2018).

**3.2 Mobility Networks and Hub Majors**

To map the ecosystem of disciplinary migration, we constructed a major-to-major transition matrix restricted exclusively to *major switches*, filtering out internal plan changes. For each ordered pair of majors (A, B), we calculated the raw volume of students moving from an origin spell in A to a destination spell in B. These flows were normalised to yield both the *feeder potential* (percentage of leavers from A targeting B) and the *attractor strength* (percentage of incoming students to B originating from A).

From this matrix, we generated a directed weighted network, where nodes represent degree majors and edges represent the volume of switching students. Node size is proportional to the total degree (sum of in-flows and out-flows). To ensure structural clarity, visualisations and centrality metrics focus on the core mobility ecosystem. Specifically, we isolated a set of high-hub majors ranked by Hub Score—defined strictly by inter-major incoming switch rates—and edges representing at least 20 student transitions. Figure 1 presents the resulting network topology.

**Figure 1.** Career Transition Network (selected majors with the highest hub score, Switches >= 20). *Nodes represent majors; edge thickness encodes the volume of students switching between them. The network reveals specific 'hub majors' acting as central attractors.*

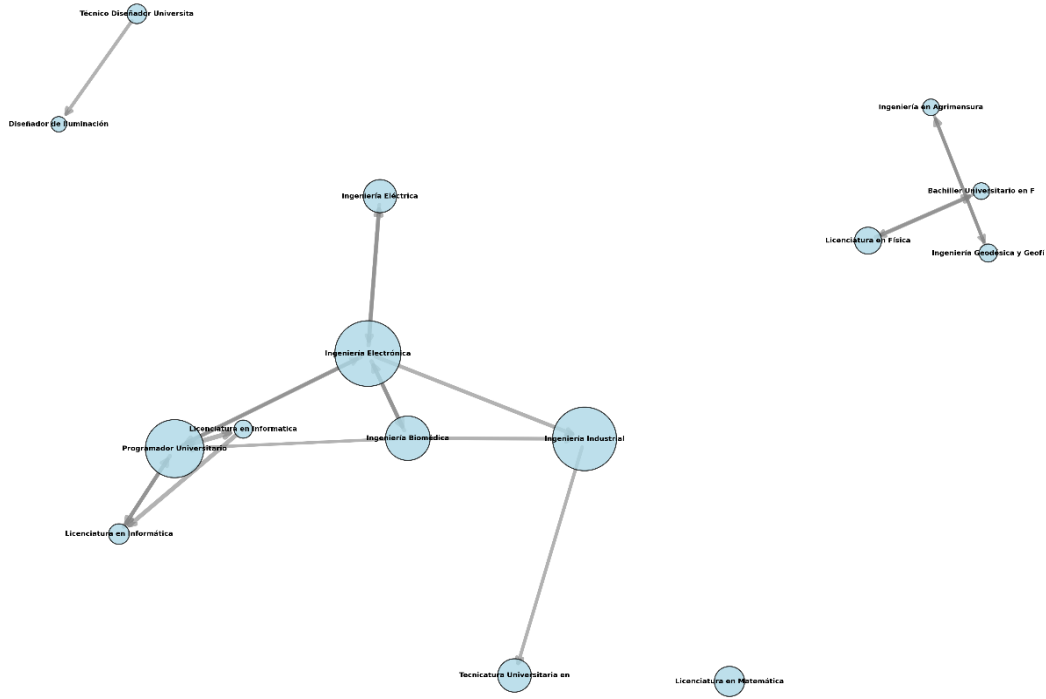

Note: Node labels are presented in the original Spanish degree titles. See Table 3 for English equivalents

Centrality metrics for these key programmes, including In-Degree, Out-Degree, and Eigenvector Centrality, are detailed in Table 3.

**Table 3.** Major Hub Metrics and Flow Statistics. *Summary of incoming and outgoing flows for selected majors with the highest hub scores.*

| major_title | n_students | n_reentry_same_plan | n_plan_change_same_title | n_switch_out | n_switch_in | reentry_rate_% | plan_change_rate_% | switch_out_rate_% | switch_in_rate_% | hub_score |
|---|---|---|---|---|---|---|---|---|---|---|
| University Technician in | 23 | 0 | 0 | 5 | 22 | 0.0 | 0.0 | 21.7 | 95.7 | 95.65 |

| Major | | | | | | | | | |
|---|---|---|---|---|---|---|---|---|---|
| Environmental Physics | | | | | | | | | |
| Licentiate in Computer Science | 383 | 13 | 0 | 156 | 310 | 7 | 0 | 85.7 | 164.1 | 164.09 |
| University Bachelor in Physics | 142 | 4 | 0 | 34 | 106 | 2.8 | 0.0 | 23.9 | 74.6 | 74.65 |
| Lighting Designer | 121 | 3 | 0 | 10 | 88 | 2.5 | 0.0 | 8.3 | 72.7 | 72.73 |
| Surveying Engineering | 140 | 4 | 0 | 25 | 73 | 2.9 | 0.0 | 17.9 | 52.1 | 52.14 |
| Sugar Engineering | 87 | 1 | 15 | 14 | 39 | 1.1 | 17.2 | 16.1 | 44.8 | 44.83 |
| University Technician in Lighting Design | 195 | 3 | 0 | 52 | 71 | 1.5 | 0.0 | 26.7 | 36.4 | 36.41 |
| Geodesic and Geophysical Engineering | 168 | 5 | 0 | 49 | 48 | 3.0 | 0.0 | 29.2 | 28.6 | 28.57 |
| University Programmer | 1723 | 43 | 0 | 283 | 482 | 2.5 | 0.0 | 16.4 | 28.0 | 27.97 |
| Electrical Engineering | 560 | 96 | 9 | 91 | 129 | 17.1 | 1.6 | 16.2 | 23.0 | 23.04 |
| University Technician in Sugar Technology and Derived Industries | 568 | 41 | 0 | 29 | 120 | 7.2 | 0.0 | 5.1 | 21.1 | 21.13 |
| Biomedical Engineering | 1006 | 323 | 0 | 170 | 156 | 32.1 | 0.0 | 16.9 | 15.5 | 15.51 |
| Licentiate in Physics | 370 | 14 | 18 | 135 | 56 | 3.8 | 4.9 | 36.5 | 15.1 | 15.14 |
| Industrial Engineering | 2069 | 406 | 0 | 242 | 298 | 19.6 | 0.0 | 11.7 | 14.4 | 14.4 |
| Electronic Engineering | 2204 | 95 | 450 | 397 | 273 | 4.3 | 20.4 | 18.0 | 12.4 | 12.39 |
| Land Surveying | 51 | 2 | 0 | 13 | 6 | 3.9 | 0.0 | 25.5 | 11.8 | 11.76 |
| Licentiate in Mathematics | 463 | 8 | 0 | 40 | 45 | 1.7 | 0.0 | 8.6 | 9.7 | 9.72 |
| Computer Engineering | 2408 | 40 | 387 | 456 | 215 | 1.7 | 16.1 | 18.9 | 8.9 | 8.93 |
| Mechanical Engineering | 1845 | 58 | 128 | 176 | 159 | 3.1 | 6.9 | 9.5 | 8.6 | 8.62 |
| Civil Engineering | 2591 | 369 | 51 | 240 | 173 | 14.2 | 2.0 | 9.3 | 6.7 | 6.68 |
| Chemical Engineering | 1823 | 60 | 191 | 279 | 109 | 3.3 | 10.5 | 15.3 | 6.0 | 5.98 |
| Electrical Engineering (BC) | 858 | 8 | 0 | 109 | 21 | 0.9 | 0.0 | 12.7 | 2.4 | 2.45 |

This network approach allows us to empirically distinguish between *hub majors* (e.g., Computing, Electronics) that act as vocational attractors and *feeder majors* (e.g., basic sciences) that distribute students across the faculty. While network analysis has been applied to labour markets (Packer et al., 2023), its application here reveals how internal curriculum structures channel student flow.

**3.3 Temporal Dynamics of Mobility Types**

Horizontal mobility is not static; it pulses with institutional rhythms. We aggregated transitions by the calendar year of the destination spell and classified them into three mutually exclusive types: (1) Major Switches (change of degree title); (2) Plan Changes (curricular update within same title); and (3) Same-Plan Re-entries (return after >= 2 years of inactivity).

This decomposition yields a 40-year timeline (1980–2019) partitioned into four structural periods (P1: 1980–1995; P2: 1996–2004; P3: 2005–2012; P4: 2013–2019). The resulting time series reveals that mobility types have distinct signatures: whereas major switching remains a constant background process, plan changes and re-entries exhibit sharp, policy-induced peaks (Figure 2).

**Figure 2.** Timeline of Academic Transitions (1980–2019). *The series highlights waves of induced mobility, particularly plan changes (orange) and re-entries (green), coincident with institutional reforms.*

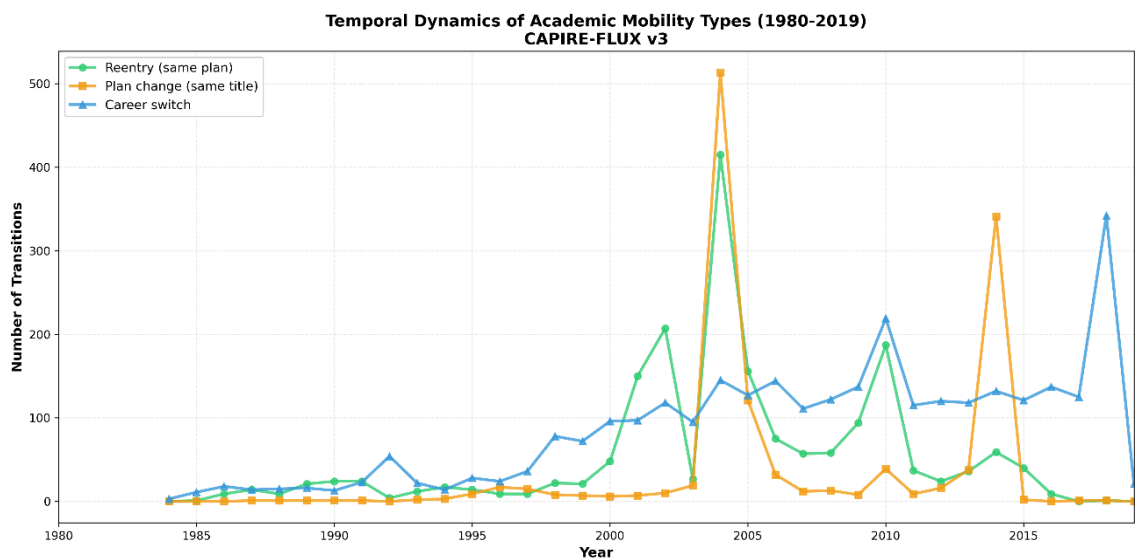

Unlike traditional survival curves (Quimio et al., 2021), this temporal decomposition foregrounds the *timing and intensity* of horizontal adjustments, demonstrating how curriculum reforms generate transient 'shockwaves' of internal migration.

**3.4 Unsupervised Learning: PCA and DBSCAN for Trajectory Archetypes**

To move from aggregate flows to student-level typologies, we constructed a compact feature set for the 17,021 students who generated active trajectory data. The feature vector included five standardised variables: total enrolment spells

(*n_spells*), binary indicators for major switching (*has_major_switch*), plan changes (*has_plan_change*), and re-entries (*has_reentry*), alongside the structural entry period (*first_period*).

We applied Principal Component Analysis (PCA) to project these profiles into a latent 2D space. The Principal Component Analysis (PCA) extracted two dominant components that together account for **61.4% of the total variance** in the background feature space. The **first component (PC1)** explains **39.7%** of the total variance and functions as a *trajectory-complexity axis*, showing an almost perfect correlation with the number of enrolment spells (r = 0.979) and substantial loadings on major switching (r = 0.720) and same-plan re-entry (r = 0.581). The **second component (PC2)** explains **21.7%** of the variance and captures temporal structure, loading strongly on entry period (r = 0.819) and inversely on re-entry (r = –0.507).

Together, these two components provide a low-dimensional representation that preserves both the ordinal nature of trajectory complexity and the historical differentiation of cohorts.

The Quantised Nature of Academic Pathways It might be argued that student mobility could be described using simple descriptive statistics. However, our PCA results reveal a fundamental structural property that traditional tables obscure: the quantised geometry of academic pathways. As shown in Figure 3, students do not distribute along a continuum; rather, they occupy distinct vertical bands along the complexity axis (PC1).

**Figure 3.** Student Trajectory Clusters in PCA Space.

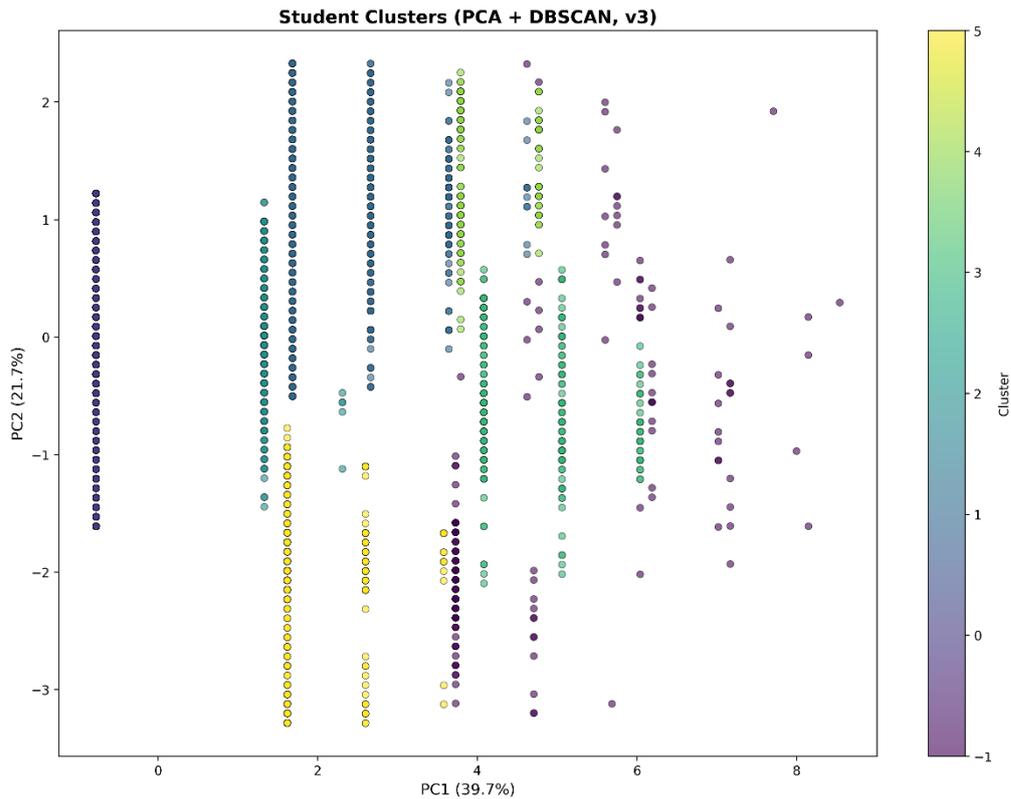

*The horizontal axis (PC1, **39.7% of total variance**) captures the stepwise increase in trajectory complexity, with strong loadings on the number of enrolment spells (r = 0.979), major switching (r = 0.720) and same-plan re-entries (r = 0.581). The vertical axis (PC2, **21.7% of total variance**) reflects cohort timing (r = 0.819 with entry period) and negatively correlates with re-entry behaviour (r = –0.507). The six DBSCAN clusters occupy discrete vertical bands, evidencing the quantised nature of academic mobility clusters.*

This banding occurs because academic transitions are discrete events—a student cannot have 1.5 spells. The near-perfect correlation between PC1 and the sum of spells and switches (r=0.967) confirms that students inhabit discrete 'quantum states' of complexity. Simple averages would smooth out these structural boundaries, hiding the specific thresholds that define student progression.

To identify archetypes within this structured space, we employed DBSCAN (Density-Based Spatial Clustering of Applications with Noise). Unlike K-Means, DBSCAN does not assume spherical clusters, allowing it to detect the naturally irregular shapes of these mobility bands (Ester et al., 1996; Hlosta et al., 2021). The algorithm identified six distinct archetypes (Stable, Switchers, Adjusters, Stable Re-entrants, Re-entrant Switchers, and Switcher-Adjusters).

**Table 4.** Cluster Summary and Characteristics. *Descriptive statistics for the six identified mobility archetypes.*

| cluster_id | cluster_name | n_students | pct_students | avg_n_spells | avg_major_switches | avg_plan_changes | pct_with_major_switch |
|---|---|---|---|---|---|---|---|
| -1 | Noise | 167 | 1.0 | 3.98 | 1.17 | 0.82 | 42.5 |
| 0 | Stable | 12474 | 73.3 | 1.0 | 0.0 | 0.0 | 0.0 |
| 1 | Switchers | 1827 | 10.7 | 2.26 | 1.26 | 0.0 | 100.0 |
| 2 | Adjusters | 925 | 5.4 | 2.01 | 0.0 | 1.01 | 0.0 |
| 3 | Re-entrant Switchers | 269 | 1.6 | 3.59 | 1.4 | 0.0 | 100.0 |
| 4 | Switcher-Adjusters | 180 | 1.1 | 3.23 | 1.19 | 1.04 | 100.0 |
| 5 | Stable Re-entrants | 1179 | 6.9 | 2.19 | 0.0 | 0.0 | 0.0 |

This combination of network analysis and unsupervised clustering provides a coherent, multi-level description of the ecosystem, connecting the macro-topology of degree offerings with the micro-structure of individual student pathways.

## 4. RESULTS

### 4.1 Spell Structure and Transition Taxonomy

The reconstructed cohort comprised **17,021 students** with valid trajectory activity between 1980 and 2019. Within this population, the majority (73.3%) followed a linear path characterised by a single continuous enrolment spell. However, a significant minority (26.7%) exhibited fragmented trajectories involving two or more spells. As illustrated in **Figure 4**, the distribution of spells per student reveals a long-tailed complexity structure; while 12,474 students completed a single spell, over 4,500 students generated multiple spells, indicating patterns of interruption, switching, or system re-entry.

Across the dataset, we identified **6,227 valid inter-spell transitions**, which were classified into three mutually exclusive structural types:

- **Major Switches (49.5%; n=3,083):** Genuine vocational re-orientation where the degree title changes.

- **Same-Plan Re-entries (30.4%; n=1,890):** Students returning to the exact same major and curriculum plan after a period of inactivity exceeding two years.

- **Plan Changes (20.1%; n=1,254):** Internal administrative migration between curriculum versions within the same degree title.

This decomposition confirms that approximately half of horizontal mobility represents vocational correction, while the remainder is driven by institutional factors (curriculum reforms) or stop-out behaviours.

**Figure 4.** Distribution of enrolment spells per student (N = 17,021). The histogram highlights the predominance of single-spell trajectories (Cluster 0 candidates) and the long tail of complex, multi-spell pathways.

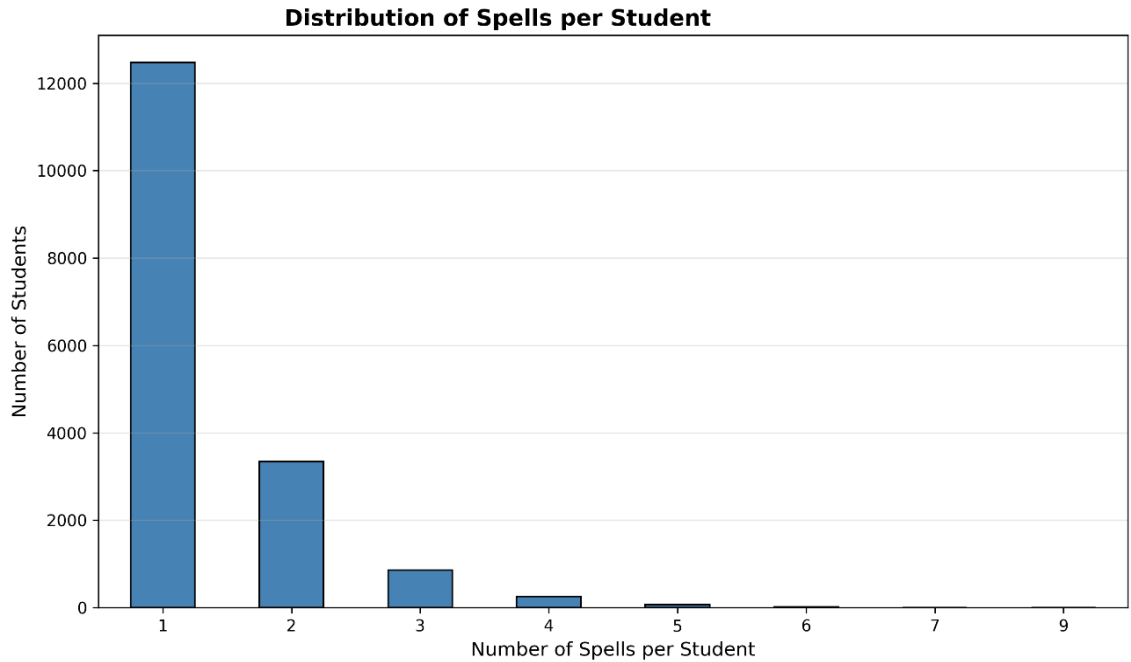

### 4.2 The Topology of Major Switching

Restricting the analysis to **major switches** reveals a highly structured ecosystem. The transition network visualized in **Figure 1** demonstrates that student flows are not random but channelled through specific corridors. The system is topologically divided into a dense **connected core** and several peripheral islands.

The core is dominated by **'hub majors'**—predominantly *Ingeniería Electrónica* (Electronic Engineering), *Ingeniería Industrial* (Industrial Engineering), and *Licenciatura en Informática* (Computer Science)—which function as central attractors (indicated by node size). The edge thickness in Figure 1 highlights distinct 'high-volume corridors', such as the flow between *Programador Universitario* and *Licenciatura en Informática*, representing a vertical articulation between technical and bachelor degrees. Conversely, majors such as *Agrimensura* (Surveying) form isolated clusters (top right), exchanging students primarily within a closed sub-system of related earth-science degrees.

### 4.3 Temporal Dynamics of Mobility Types

Horizontal mobility is not static; it pulses with institutional rhythms. The temporal decomposition in **Figure 2** reveals that the three mobility types exhibit distinct signatures over the 40-year window:

1. **Plan Changes (Orange Line):** These exhibit violent spikes coincident with curriculum reforms, most notably the massive wave of administrative migration in 2004 and a secondary wave in 2014. These are induced transitions, reflecting policy implementation rather than student choice.

2. **Re-entries (Green Line):** Returns to the same plan peaked significantly in the early 2000s (reaching >400 events/year) but show a marked decline after 2010. This suggests a shift in student behaviour or institutional retention capability, with fewer students successfully returning after long gaps in recent cohorts.

3. **Major Switches (Blue Line):** In contrast to the other types, career switching shows a steady, secular upward trend, acting as a constant 'background radiation' of vocational adjustment that grows proportionally with faculty enrolment.

**4.4 Hub Metrics: Distinguishing Mobility Profiles**

To understand the functional role of different degrees, we decomposed the incoming flow of the top-15 majors. **Figure 5** classifies majors by the nature of their 'churn'.

- **Attractor Hubs:** Degrees like *Licenciatura en Informática* and *Diseñador de Iluminación* are dominated by **Incoming Switches** (blue bars), indicating they serve as destinations for students re-orienting from other fields.

- **Re-entry Clusters:** Majors such as *Ingeniería Biomédica* and *Ingeniería Industrial* show significant proportions of **Same-Plan Re-entries** (green bars), suggesting these demanding programmes experience high rates of 'stop-out' behaviour where students pause and later return.

- **Reform-Sensitive Majors:** *Ingeniería Electrónica* displays a notable component of **Plan Changes** (orange bars), reflecting the impact of specific curricular updates on its student body.

**Figure 5.** Mobility profile by major (Top-15 by Hub Score). The stacked bars distinguish between incoming career switchers (blue), internal plan changers (orange), and returning students (green), revealing the distinct functional role of each major in the ecosystem.

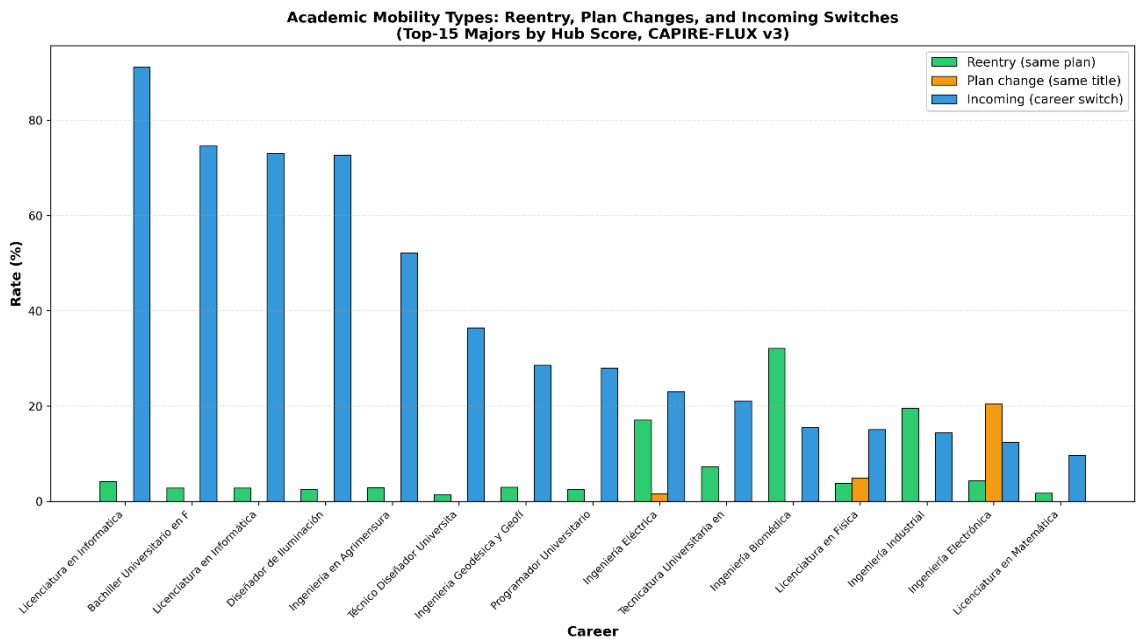

### 4.5 Student Mobility Archetypes and the Vertical Band Phenomenon

The application of PCA and DBSCAN to the student feature set (N=17,021) reveals that academic trajectories are structurally **quantised**. As demonstrated in **Figure 3**, students do not distribute along a continuum of complexity; rather, they occupy discrete **vertical bands** along the first principal component (PC1).

The PCA embedding confirms that academic trajectories are not continuously distributed but instead form discrete, quantised bands of complexity. The first principal component (**PC1, 39.7% of total variance**) is dominated by the number of enrolment spells (r = 0.979) and by major-switch behaviour (r = 0.720), effectively capturing an ordinal "complexity axis". The second component (**PC2, 21.7% of total variance**) reflects temporal and re-entry structure, with strong loadings on entry period (r = 0.819) and a negative association with same-plan re-entries (r = –0.507). In combination, these two components retain **61.4%** of the total variance, enough to reveal the striking vertical-band geometry of the student-trajectory space. Within this structured space, DBSCAN identified six robust archetypes:

1. **Cluster 0: Estables (Stable, 73.3%):** Occupying the first vertical band (leftmost dark purple), these students follow linear, single-spell trajectories.

2. **Cluster 1: Switchers (10.7%):** Located in the second and third bands, these students change major exactly once without re-entry, representing pure vocational correction.

3. **Cluster 5: Stable Re-entrants (6.9%):** A critical finding, these students (visible as specific density groups in the multi-spell bands) return to the *same* major after a gap. Notably, this archetype is prevalent in 1990s–2000s cohorts but nearly absent in the 2010s, confirming the temporal decline observed in Figure 2.

4. **Complex Clusters (Adjusters/Mobile Re-entrants):** The remaining clusters capture high-entropy trajectories involving multiple switches and reforms, occupying the sparse right-hand side of the projection.

## 5. DISCUSSION

### 5.1 From Binary Outcomes to Quantised Mobility

The results demonstrate that horizontal mobility within a large engineering and science faculty is neither random noise nor a marginal phenomenon, but a structurally organised dimension of the academic ecosystem. Almost half of all inter-spell transitions correspond to major switches, and more than a quarter of the student population experiences multiple enrolment spells. Yet, traditional retention models continue to compress this rich diversity into binary outcomes—'retained' versus 'dropout'—or coarse, aggregate measures of time-to-degree.

By applying the CAPIRE framework, we offer a fundamental reframing: student progression is not a continuous gradient of success but a **quantised geometric structure**. The vertical bands observed in the PCA projection (Figure 3) provide empirical evidence that trajectory complexity advances in discrete 'quantum' steps. This is mathematically corroborated by the near-perfect correlation ($r = 0.979$) between the first principal component and the count of administrative events.

This finding has profound methodological implications for Higher Education research. Statistical models that treat trajectory complexity as a continuous variable (e.g., standard linear regression on semester counts) inevitably under-represent the critical importance of crossing discrete thresholds—such as the qualitative leap from a single-spell trajectory to a multi-spell, multi-major pathway. Our analysis suggests that modelling frameworks based on discrete states, such as Multi-State Survival Models or Sequence Analysis, are not merely alternative options but are conceptually required to capture the ordinal nature of these pathways.

### 5.2 Hub Majors and Systemic Fragility

The topology of the transition network reveals that the faculty functions as a complex system with a 'core-periphery' structure. A small set of **'hub majors'**—specifically *Ingeniería Electrónica*, *Ingeniería Industrial*, and *Licenciatura en Informática*—absorb the vast majority of incoming switchers. These programmes combine high incoming volumes with relatively modest re-entry rates, indicating that they function as 'attractor nodes' for vocational re-orientation rather than as fallback options for failing students.

From a complexity science perspective, this concentration of flow introduces a dimension of **systemic risk** often overlooked in faculty planning. These hubs provide a critical stabilising function, offering viable pathways for students whose initial choices (often in basic sciences or rigid traditional degrees) prove misaligned with their interests. However, this high centrality also implies system-wide fragility. Radical curriculum reforms, capacity caps, or changes in admission criteria within a 'super-hub' like *Ingeniería Electrónica* could trigger cascade effects, blocking the primary corridors of internal migration and forcing students out of the system entirely. Thus, the management of hub majors is not solely a departmental concern but a matter of ecosystem resilience.

**5.3 The Extinction of the 'Stable Re-entrant' and Institutional Rigidity**

Perhaps the most striking finding of the CAPIRE v3 analysis is the temporal collapse of **Cluster 5: Stable Re-entrants**. This archetype, defined by students who withdraw for significant periods (gap years or stop-outs) and successfully re-enter the same major, was a dominant feature of the 1990s and early 2000s cohorts, accounting for 86.3% of the cluster's population. In stark contrast, this behaviour has become statistically negligible in the post-2010 cohorts, representing only 1.4% of cases.

This 'extinction' suggests a fundamental shift in the institutional interaction between students and the university. While curriculum reforms (visible as spikes in the 'Plan Change' timeline) have modernised content, the decline in same-plan re-entries implies that the system has become effectively more rigid. The "revolving door" that once allowed students to pause their studies due to economic or personal crises and return years later appears to have closed, likely due to stricter re-admission regulations or the accelerated obsolescence of credit in modernised curricula.

While standard indicators might interpret this as an improvement in 'efficiency' (fewer students lingering in the system), it may also signal a loss of inclusivity. The contemporary university ecosystem appears optimised for high-velocity, linear trajectories, effectively selecting against non-linear life-courses and potentially excluding a demographic that requires temporal flexibility to succeed.

**5.4 Archetypes as Priors for Agent-Based Simulation**

The six mobility archetypes identified via DBSCAN offer more than a static typology; they represent distinct behavioural agents. While numerical dominance resides with the **Stable** cluster (73.3%), the dynamism of the system is driven by the **Switchers** (10.7%), **Adjusters** (5.4%), and the dwindling **Re-entrants**.

For institutional practice, this suggests moving beyond generic "at-risk" flags towards archetype-specific support strategies:

- **For Switchers:** Early advising and transparent 'gateway' courses could reduce the friction of vocational exploration, converting a potential dropout into an efficient transfer.
- **For Adjusters:** The identification of this cluster proves that curriculum reforms impose a hidden administrative tax on students. Monitoring 'bottleneck' courses during transition periods is essential to prevent these adjustments from becoming attrition events.

Furthermore, from a modelling perspective, these archetypes provide the necessary empirical calibration for **Agent-Based Models (ABM)** of higher education systems. By parametrising agents with the transition probabilities and discrete state-behaviours observed in this study (e.g., an agent's probability of becoming a 'Switcher' vs. a 'Stable Re-entrant'), institutions can move towards developing **Digital Twins** of their academic population. Such simulations would allow administrators to test counterfactual scenarios—for example, *"What happens to system-wide retention if we modify the re-entry rules for the Electronics hub?"*—before implementing potentially disruptive policies in the real world.

## 6. LIMITATIONS AND FUTURE WORK

### 6.1 Constraints of the Current Study

While this study leverages a robust 40-year longitudinal dataset, several limitations must be acknowledged. Firstly, the analysis relies on administrative data from a single large faculty within the Argentine public university system. While the sheer volume of records (N = 24,016) allows for high-resolution internal mapping, the specific topology of the 'hub majors' (e.g., the centrality of *Ingeniería Electrónica*) is undoubtedly context-dependent. However, the qualitative dynamics observed—specifically the friction generated by rigid prerequisite chains and the 'shockwaves' of curriculum reform—are likely to generalise well to other engineering-intensive institutions in Latin America and Europe that share similar degree structures.

Secondly, our taxonomy of mobility is operationally defined by administrative codes and temporal thresholds. The distinction between a 'stop-out' (temporary withdrawal) and a 'dropout' (permanent departure) relies on our two-year inactivity threshold. Sensitivity analyses suggest the main archetypes are robust, but alternative definitions (e.g., varying the gap window from 1 to 5 years) could yield more granular sub-types of re-entry behaviour. Furthermore, as the data is purely administrative, it lacks socio-economic or psychometric variables. We observe *how* students move, but we cannot causally test *why* a specific student chooses to switch majors versus leaving the system entirely.

**6.2 Future Avenues: Towards Agent-Based Simulation**

The structural and descriptive nature of CAPIRE provides the necessary foundation for the next generation of predictive modelling. We propose three specific avenues for future research:

1. **From Description to Causal Inference:** Future work should integrate this mobility framework with causal inference methods (such as Difference-in-Differences or Regression Discontinuity Designs) to evaluate the impact of specific policy interventions. For instance, the sharp peaks in 'Plan Changes' observed in 2004 and 2014 provide natural experiments to estimate the causal effect of forced curriculum migration on subsequent time-to-degree.

2. **Calibrating Agent-Based Models (ABM):** Perhaps the most promising extension of this work is the development of a **Digital Twin** of the faculty. The transition matrices and mobility archetypes identified here serve as empirical priors for calibrating agent-based simulations. By parameterising agents with the discrete state-transition probabilities observed in our network analysis, researchers could simulate counterfactual scenarios—such as the removal of a 'hub major' or the relaxation of re-entry rules—to predict emergent systemic risks before they manifest in reality.

3. **Predictive Modelling with Sequence Analysis:** While PCA and DBSCAN successfully identified structural archetypes, future research could employ deep learning techniques (e.g., LSTM networks or Transformers) on the raw event sequences to predict individual student trajectories in real-time. This would move the field from *post-hoc* classification to proactive early warning systems.

Despite these limitations, the CAPIRE framework demonstrates that internal mobility systems in higher education can be mapped, decomposed, and quantified with sufficient resolution to inform institutional decision-making. By treating horizontal mobility as a first-class object of analysis, institutions can gain a realistic understanding of how students traverse complex degree ecosystems.

**7. CONCLUSIONS**

**7.1 Mobility as a Structural, Quantised Phenomenon**

This study challenges the prevailing binary view of student success, demonstrating that horizontal mobility is not random noise surrounding retention metrics but a structural dimension of how students inhabit the institutional ecosystem. By reconstructing 40 years of administrative history, we have shown that apparently chaotic pathways resolve into a small set of recurrent, stable configurations.

The primary theoretical contribution of this work is the empirical demonstration of the **quantised geometry** of academic pathways. As evidenced by the vertical banding in our PCA results, student trajectories do not vary along a continuum but occupy discrete complexity states. A student is either stable, a single-switcher, or a multi-spell re-entrant; there are no intermediate states. This implies that student progression in rigid engineering/science curricula is shaped as much by the discrete 'quantum' rules of institutional design as by individual agency.

### 7.2 The Efficiency Paradox and the Extinction of Flexibility

Empirically, the most significant—and potentially alarming—finding is the near-total extinction of the 'Stable Re-entrant' archetype (Cluster 5) in recent cohorts. In the 1990s and 2000s, the system accommodated a significant population of students who paused their studies for years and successfully returned. In the post-2010 era, this behaviour has vanished (<1.5 % of cases). This suggests that while modern curricula may be more efficient in terms of graduation rates, the system has lost its **temporal flexibility**, effectively filtering out non-linear life courses. This 'efficiency paradox' raises urgent questions about inclusivity in the modern university: have we engineered a system that is faster but less forgiving?

### 7.3 Systemic Resilience and Hub Management

Our network analysis highlights that the faculty behaves as a complex adaptive system dependent on critical nodes. The identification of 'Super-Hubs' in Computing and Electronics reveals that these majors perform a systemic function beyond their disciplinary remit: they act as the primary vocational correctors for the entire faculty. Consequently, the stability of these programmes is a matter of ecosystem resilience. Institutional planning must recognise that capacity constraints or failures in these hubs will propagate cascade effects throughout the entire transition network.

### 7.4 CAPIRE as a Blueprint for Institutional Intelligence

Finally, CAPIRE provides a replicable, modular blueprint for other universities to reconstruct this structure from their own routine data. Its core design principles—using gap-based spells, distinguishing induced plan changes from vocational switches, and treating archetypes as discrete states—are portable across institutions.

For engineering schools and professional faculties, the implications are direct. Monitoring dashboards must evolve beyond 'retention rates' to track 'mobility flows'.

By aligning analytics with the discrete, stepwise nature of educational pathways, institutions can transform what used to be treated as administrative noise into usable intelligence for policy design, moving from a paradigm of **retention management** to one of **pathway orchestration**.